\RequirePackage[T1]{fontenc}
\documentclass[journal]{IEEEtran}
\usepackage{graphicx}
\usepackage{booktabs}
\usepackage[table]{xcolor}
\definecolor{CRFNetShade}{RGB}{223,240,255}
\usepackage{amsmath}
\usepackage{amssymb}
\usepackage{array}
\usepackage{pifont}
\usepackage{multirow}
\usepackage{hyperref}
\hypersetup{
    colorlinks=true,
    linkcolor=blue,
    urlcolor=blue,
    citecolor=blue}

\begin{document}

\title{Cross-Layer Refinement and Frequency-aware Fusion Network for Multi-Source Remote Sensing Image Classification}

\author{Wenhan Xiao,  
Chuanzheng Gong,
Feng Gao, \emph{Member, IEEE},
Junyu Dong, \emph{Member, IEEE}, \\
Qian Du, \emph{Fellow, IEEE}

\thanks{
 This work was supported in part by National Key R \& D Program of China under Grant 2024YFF0908102, Key R \& D Program of Shandong Province under Grant 2025CXPT185, and in part by the Natural Science Foundation of Shandong Province under Grant ZR2024MF020. (\textit{Corresponding author: Feng Gao})

 Wenhan Xiao, Chuanzheng Gong, Feng Gao, and Junyu Dong are with the State Key Laboratory of Physical Oceanography, Ocean University of China, Qingdao 266100, China. (Email: gaofeng@ouc.edu.cn)

 Qian Du is with the Department of Electrical and Computer Engineering, Mississippi State University, Starkville, MS 39762 USA.
}

}

\markboth{Under review at IEEE Geoscience and Remote Sensing Letters}{}

\maketitle

\begin{abstract}
Multi-source remote sensing image classification benefits from complementary spectral and structural information. However, identifying discriminative cross-modal features remains challenging. This letter proposes a Cross-Layer Refinement and Frequency-aware Fusion Network (CRFNet) within a lightweight dual-branch architecture. Specifically, a \textit{Cross-Layer Refinement Attention (CLRA)} module propagates selected attention relationships across successive layers and employs top relation selection to progressively strengthen discriminative spatial dependencies while suppressing irrelevant responses. Furthermore, a \textit{Frequency-aware Complementary Fusion (FCF)} module combines frequency-wise interactions with cross-modal channel attention in the Fourier domain. The resulting representations are then converted into spatial-domain gates to selectively modulate complementary features. Experiments on the Houston 2013 and Augsburg datasets demonstrate that the proposed CRFNet consistently outperforms several state-of-the-art methods. The codes are publicly available at \url{https://github.com/oucailab/CRFNet}.

\end{abstract}


\IEEEpeerreviewmaketitle

\section{Introduction}


\IEEEPARstart{M}{ulti-source} remote sensing data classification combines observations of the same scene from heterogeneous sensors \cite{song26scis}. By integrating complementary information acquired from different sensors, it provides a more comprehensive characterization of land-cover properties than relying on a single data source. Among these data sources, hyperspectral images (HSIs) provide abundant spectral information, and the detailed spectral signatures enable the discrimination of land-cover categories with similar visual appearances \cite{li25enhance}. Hence, HSI is particularly valuable for fine-grained classification \cite{chu26ugmft}. 


However, HSI commonly suffers from limited spatial resolution and high spectral redundancy, which may hinder the accurate characterization of spatial structures. Therefore, integrating HSI with complementary data, such as synthetic aperture radar (SAR) or light detection and ranging (LiDAR), offers an effective way to jointly exploit spectral, structural, and elevation information \cite{gcz26tgrs}. The all-weather, day-and-night imaging capability of SAR data enables reliable observations under cloud cover or poor illumination \cite{duan26tip}. LiDAR data captures accurate elevation and three-dimensional geometric information \cite{zhao26sgf}. Motivated by these complementary properties, this letter investigates the fusion of HSI with LiDAR or SAR data for land-cover classification.


Accordingly, the fusion of multi-source data has attracted considerable attention in land-cover classification. Existing methods generally integrate heterogeneous observations at the data, feature, or decision level \cite{lc25tgrs}. Among these strategies, feature-level fusion provides a flexible means of combining complementary representations learned from different modalities. Bi et al. \cite{bi26fmtunet} employ a synergistic Mamba fusion module to dynamically align and integrate optical and elevation features through cross-modal spatial interactions at the encoder level. Gong et al. \cite{gcz26tgrs} propose RSCNet, which selects task-relevant HSI bands using cross-source guidance and adaptively fuses them with SAR or LiDAR features through cross-source attention and local--global refinement. Zhang et al. \cite{zhang26if} progressively fuse multi-source features through multi-scale alignment and adaptively gated cross-modal attention, combining the outputs of all stages using learnable fusion weights. Effective fusion remains challenging because different sensors exhibit substantial discrepancies in imaging mechanisms and spatial characteristics. 

\begin{figure*}[tbp]
\centering
\includegraphics[width=0.9\textwidth]{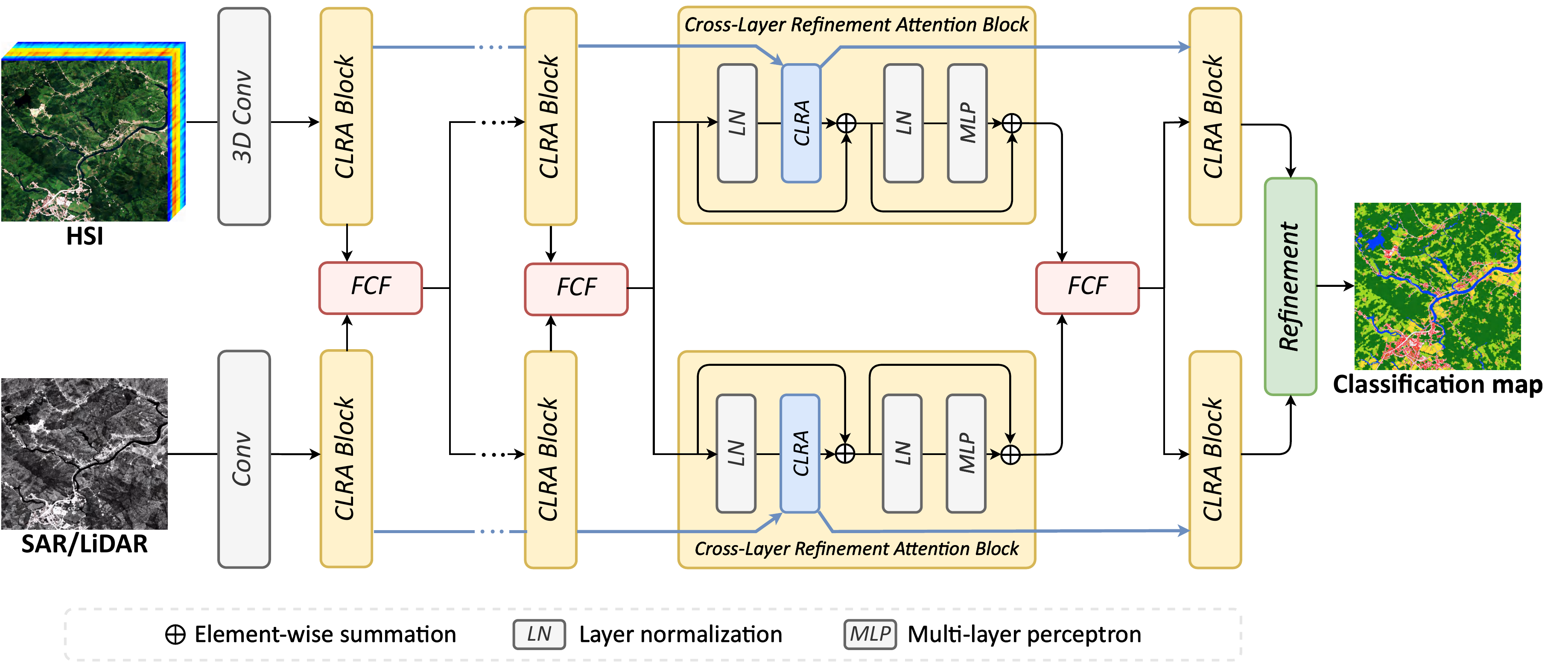}
\caption{Overview of the proposed Cross-Layer Refinement and Frequency-aware Fusion Network (CRFNet) for multi-source remote sensing classification. The dual branches extract features from HSI and SAR/LiDAR data, respectively. Cross-Layer Refinement Attention (CLRA) blocks progressively refine features through cross-layer interactions, while Frequency-aware Complementary Fusion (FCF) modules integrate complementary information across modalities at successive stages.}
\label{fig-frame}
\end{figure*}

Despite these promising advances, two key challenges persist in multi-source remote sensing data classification: \ding{202} \textbf{How to progressively identify and strengthen discriminative features.} In multi-source feature extraction, complementary cues are often intertwined with noise and background interference, making it difficult to isolate features that reliably distinguish land-cover classes. This challenge calls for a progressive refinement mechanism that preserves informative features and increasingly focuses feature aggregation on task-relevant cues. \ding{203} \textbf{How to jointly reinforce shared cross-modal features and preserve modality-specific complementary cues.} Most existing fusion strategies that primarily emphasize cross-modal similarity may suppress useful differential features, while indiscriminate aggregation can introduce irrelevant information. The challenge therefore lies in selectively retaining complementary differential cues that enhance class discrimination.

To address these challenges, we propose \textbf{C}ross-Layer \textbf{R}efinement and \textbf{F}requency-aware Fusion \textbf{Net}work (\textbf{CRFNet}) for multi-source remote sensing classification. Specifically, we introduce \textit{Cross-Layer Refinement Attention (CLRA)}, which propagates retained attention relations across successive layers to guide feature refinement. By combining cross-layer guidance with top relation selection, CLRA progressively strengthens discriminative associations while suppressing irrelevant responses. Furthermore, we develop a \textit{Frequency-aware Complementary Fusion (FCF)}  module to facilitate cross-modal interactions in the Fourier domain. FCF combines frequency interactions with cross-modal channel attention and transforms the aggregated representation into a spatial-domain gate for feature modulation. 

The contributions of this letter are summarized as follows:

\begin{itemize}

\item We design CLRA that propagates retained attention relations across successive layers. Cross-layer guidance and layer-wise top-$k$ selection progressively strengthen discriminative associations.

\item We develop FCF module that combines frequency interactions with cross-modal channel attention. The obtained features are transformed into a spatial-domain gate to selectively enhance cross-modal information.

\item Extensive experiments on two benchmark datasets demonstrate that the proposed CRFNet outperforms several state-of-the-art methods.

\end{itemize}

\section{Methodology}

\subsection{Overall Framework of the Proposed CRFNet}

As illustrated in Fig. \ref{fig-frame}, the proposed CRFNet adopts a dual-branch architecture for multi-source remote sensing classification. A 3D convolution first extracts spectral--spatial features from the HSI input, while a convolutional layer encodes structural information from the structural SAR/LiDAR data. Both branches subsequently employ stacked Cross-Layer Refinement Attention (CLRA) blocks to progressively refine their representations through interactions between features from different depths. Each CLRA block comprises a cross-layer attention unit and a multi-layer perceptron (MLP), with layer normalization and residual connections. At successive stages, Frequency-aware Complementary Fusion (FCF) modules facilitate cross-modal information exchange by reinforcing shared features and retaining modality-specific complementary cues. The resulting features are fed into subsequent refinement module to produce the land-cover classification map.

\subsection{Cross-Layer Refinement Attention}

\begin{figure}[tbp]
\centering
\includegraphics[width=3.3in]{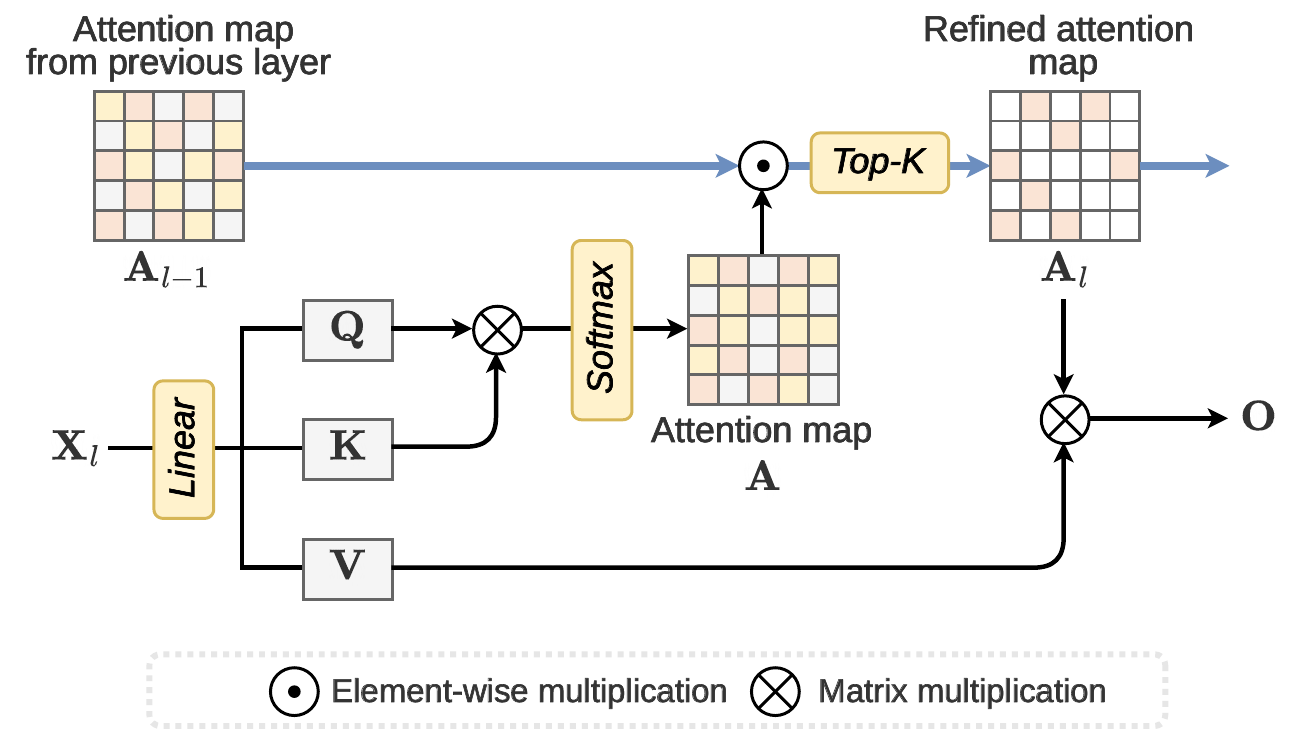}
\caption{Illustration of the Cross-Layer Refinement Attention (CLRA).}
\label{fig-clra}
\end{figure}

In multi-source feature extraction, complementary cues
are often intertwined with noise and background interference,
making it difficult to isolate features that reliably distinguish land-cover classes. To address this issue, we design the Cross-Layer Refinement Attention (CLRA) to use attention information retained from the previous layer to guide relation modeling in the current layer, and then selects the dominant features to restrict the propagation of irrelevant information. The same spatial CLRA design is applied independently in the HSI and structural branches; their attention masks are not shared. Therefore, CLRA can focus feature aggregation on complementary cues that contribute to class discrimination.

\textbf{Current-Layer Attention Computation.} The input feature tensor at $l$-th layer is $\mathbf X_l \in\mathbb R^{R\times R\times C}$, where $R$ and $C$ denote the input patch size, and number of channels, respectively. A linear layer is employed to the generate query $\mathbf{Q}$, key $\mathbf{K}$, and value $\mathbf{V}$. The similarities between spatial positions are first calculated and normalized using Softmax over the key positions:

\begin{equation}
\widetilde{\mathbf A}_l=\mathrm{Softmax}
\left(
\frac{\mathbf Q \mathbf K ^\top}{\sqrt d}
\right).
\end{equation}
This step establishes global spatial relationships within the current layer but does not yet incorporate the selection information from the previous layer.

\textbf{Cross-Layer Attention Refinement.} The refined attention propagated from the previous layer is $\mathbf{A}_{l-1}$. To introduce cross-layer guidance, the current attention map $\widetilde{\mathbf A}_l$ is multiplied element-wise by the refined attention map $\mathbf{A}_{l-1}$ propagated from the previous layer:
\begin{equation}
  \mathbf B_l=\widetilde{\mathbf A}_l\odot\mathbf A_{l-1},
\end{equation}
where $\odot$ denotes element-wise multiplication. At the first layer, $\mathbf B_1=\widetilde{\mathbf A}_1$. This operation uses previously retained relations to modulate current attention responses, assigning higher selection priority to associations supported by both layers and attenuating those receiving weak support from the previous layer.

\textbf{Top Relation Selection.} We compute the refined attention map by applying sparse selection to $\mathbf{B}_l$:
\begin{equation}
  \mathbf A_l=\mathcal{T}(\mathbf B),
\end{equation}
where $\mathcal{T}(\cdot)$ is the learnable top-$k$ selection operator:

\begin{equation}
[\mathcal{T}(\mathbf{B})]_{ij}
=
\begin{cases}
\mathbf{B}_{ij}, & \mathbf{B}_{ij} \in \text{top-}k( j), \\
0, & \text{otherwise}.
\end{cases}
\end{equation}
An adaptive selection of the top-$k$ contribution scores are employed to preserve the most dominant components and remove the useless ones. This dynamic selection makes the attention from dense to sparse. The top-$k$ selection ratio $\alpha$ is an important parameter, and its effect will be analyzed in the experiments.

The refined attention map $\mathbf A_l$ is combined with $\mathbf{V}$ to generate the output:
\begin{equation}
  \mathbf{O} = \mathbf{A}_l \otimes \mathbf{V}.
\end{equation}
Meanwhile, $\mathbf{A}_l$ is propagated to the next layer to guide subsequent attention refinement, enabling successive layers to build upon the retained associations and progressively focus feature aggregation on discriminative cues.

\subsection{Frequency-aware Complementary Fusion}

\begin{figure}[]
\centering
\includegraphics[width=3.4in]{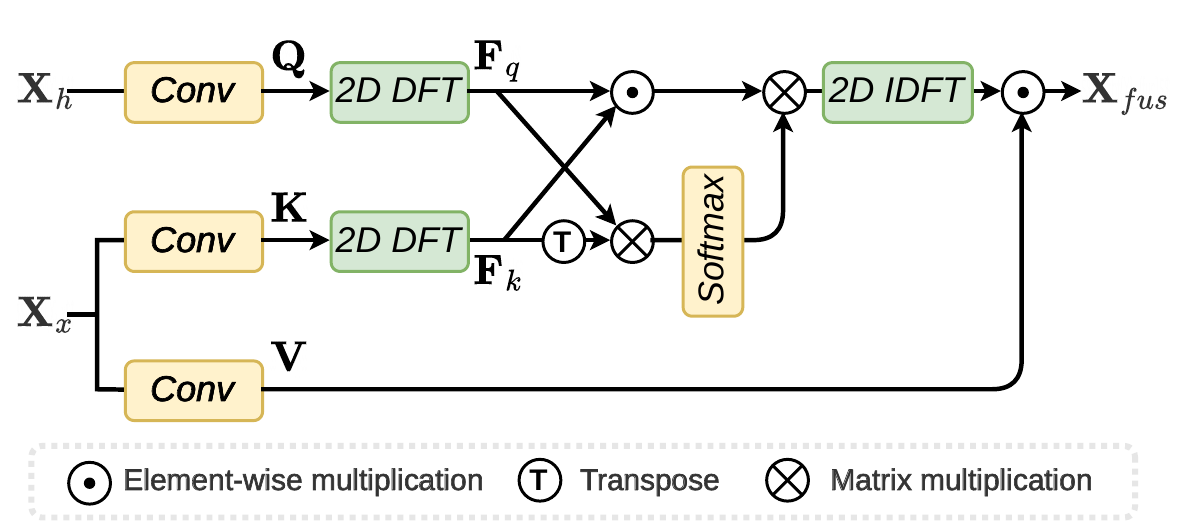}
\caption{Illustration of the Frequency-aware Complementary Fusion (FCF) module.}
\label{fig-fcf}
\end{figure}

In multi-source data fusion, most existing fusion models emphasize cross-modal similarity may suppress useful differential features, while indiscriminate aggregation can introduce irrelevant information. To this end, we propose Frequency-aware Complementary Fusion (FCF) module for multi-source feature fusion. As shown in Fig. \ref{fig-fcf}, given the hyperspectral features $\mathbf F_h$ and LiDAR or SAR features $\mathbf F_x$, three linear projections generate the query $\mathbf Q$, key $\mathbf K$, and value $\mathbf V$ representations as:
\begin{equation}
\mathbf Q=\phi_q(\mathbf F_h), ~~
\mathbf K=\phi_k(\mathbf F_x), ~~
\mathbf V=\phi_v(\mathbf F_x),
\end{equation}
where $\phi_q$, $\phi_k$ and $\phi_v$ denote independent linear mappings. The projected features share the same dimensions $C\times H\times W$. We apply two-dimensional discrete Fourier transform (DFT) to $\mathbf Q$ and $\mathbf K$, followed by spatial flattening:
\begin{equation}
\mathbf F_q=\mathcal F(\mathbf Q),\qquad
\mathbf F_k=\mathcal F(\mathbf K),
\end{equation}
where $\mathcal{F}$ denotes the Fourier transform. Two parallel paths subsequently capture complementary cross-modal interactions. The first computes an element-wise frequency interaction:
\begin{equation}
  \mathbf P=\mathbf F_q\odot\mathbf F_k,
\end{equation}
which retains interactions at corresponding channels and frequency locations. The second constructs a cross-modal channel attention matrix:
\begin{equation}
\mathbf A=
\mathrm{Softmax}\left(
\frac{\mathbf F_q\mathbf F_k^{T}}
{\sqrt d}
\right),
\end{equation}
where Softmax is applied along each row, yielding $\mathbf A\in\mathbb R^{C\times C}$. The channel attention then aggregates the frequency interaction features:
\begin{equation}
\mathbf Z=\mathbf A\mathbf P,
\qquad
\mathbf Z\in\mathbb R^{C\times M}.  
\end{equation}

Finally, the aggregated representation is reshaped and transformed back into the spatial domain to modulate the value features:
\begin{equation}
\mathbf G= \mathcal F^{-1}\left(\mathbf Z\right),
\qquad
\mathbf F_{fus}=\mathbf G\odot\mathbf V.
\end{equation}
Here, $\mathcal F^{-1}$ denotes the inverse Fourier transform. By combining frequency-wise interactions with cross-channel aggregation, FCF produces a joint cross-modal response that adaptively modulates structural information for subsequent classification.

\section{Experimental Results and Analysis}

\subsection{Datasets and Settings}

We evaluate the performance of the proposed CRFNet on two benchmark datasets for multi-source remote sensing data classification. The first is the Houston 2013 dataset, released for the 2013 IEEE GRSS Data Fusion Contest. It covers the University of Houston campus and its surrounding urban areas, USA, and comprises an HSI and a co-registered LiDAR-derived digital surface model. Both modalities contain $349 \times 1905$ pixels at a spatial resolution of 2.5 m. The second is the Augsburg dataset \cite{hong21mul}, acquired over Augsburg, Germany, from which we use the co-registered HSI and SAR observations. The scene contains $332 \times 485$ pixels at a spatial resolution of 30 m and includes seven land-cover classes. Its HSI comprises 180 spectral bands covering 0.4--2.5 $\mu$m, while the SAR data provide complementary microwave scattering information. 

Our experiments were conducted on an NVIDIA GeForce RTX 4090 GPU. It is trained for 200 epochs using the AdamW optimizer, with a batch size of 128, a learning rate of $1\times10^{-4}$, and an embedding dimension of $C=64$. To evaluate the performance of the proposed CRFNet, we compare it against other state-of-the-art methods, including S$^2$ENet \cite{fang22s2enet}, DFINet \cite{gao22dfinet}, AsyFF \cite{li22asyffnet}, ExViT \cite{yao23exvit}, HCT \cite{zhao23hct}, MACN \cite{li23macn}, and RSCNet \cite{gcz26tgrs}. To quantitatively evaluate the experimental results, we employ three metrics: Overall Accuracy (OA), Average Accuracy (AA), and Kappa coefficient. OA and AA are reported as percentages, and Kappa is scaled by 100 in the tables. Accuracy differences are expressed in percentage points. 

\subsection{Parameter Analysis}

\begin{figure}[!t]
\centering
\includegraphics[width=\linewidth]{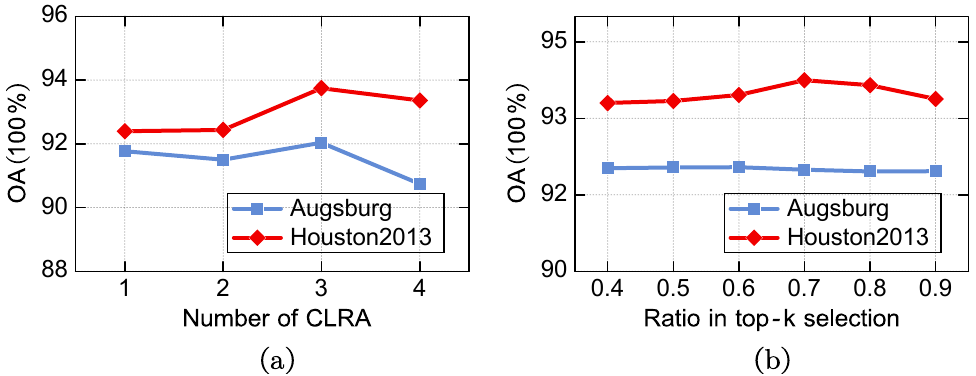}
\caption{Parameter analysis of CRFNet. (a) Number of CLRA blocks. (b) Ratio in top-$k$ selection. }
\label{fig-param}
\end{figure}

\textbf{Number of CLRA Blocks.} We investigate the effect of the number of CLRA blocks by varying the block number from 1 to 4. As shown in Fig. \ref{fig-param}(a), both datasets achieve their highest OA with three blocks. Adding a fourth block reduces OA, particularly on Augsburg. These results suggest that progressively refining attention benefits feature discrimination, but additional network depth does not necessarily yield further improvements. Therefore, we use three CLRA blocks in our experiments.

\textbf{Ratio in Top-$k$ Selection.} We examine the influence of the top-$k$ selection ratio by varying it from 0.4 to 0.9. As shown in Fig. \ref{fig-param}(b), the OA on Houston 2013 initially increases, peaks at a ratio of 0.7, and then decreases as the ratio increases further. This suggests that an appropriate selection ratio balances retaining informative attention contributions and suppressing irrelevant ones. In contrast, the performance on Augsburg remains relatively stable across the tested ratios, indicating lower sensitivity to this hyperparameter.

\subsection{Classification Performance Comparison}

\begin{figure}[!t]
\centering
\includegraphics[width=0.98\linewidth]{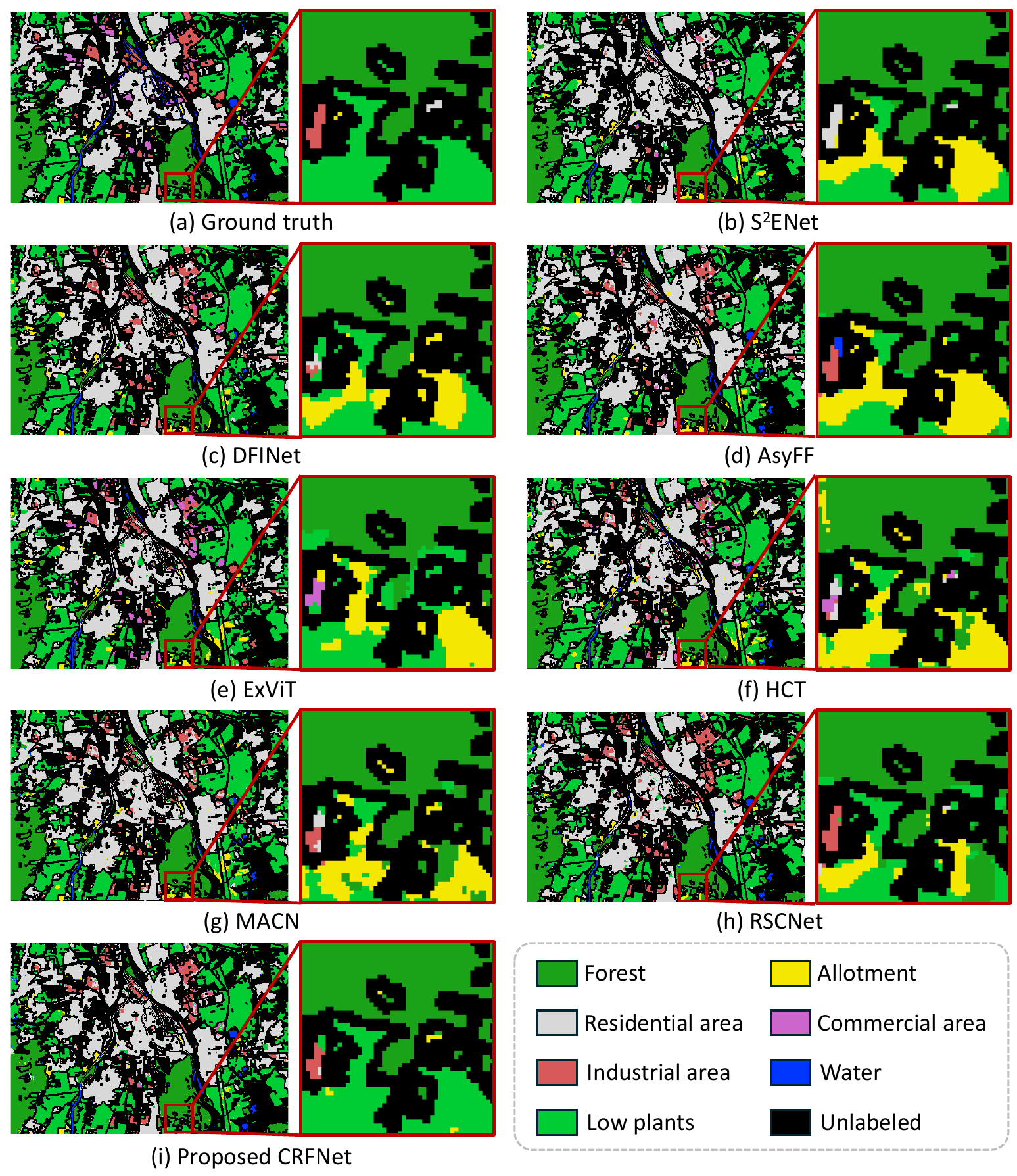}
\caption{Qualitative comparison of the classification maps on the Augsburg dataset, with enlarged views of a selected region. (a) Ground truth. (b) S$^2$ENet. (c) DFINet. (d) AsyFF. (e) ExViT. (f) HCT. (g) MACN. (h) RSCNet. (i) Proposed CRFNet.}
\label{fig:augsburg-res}
\end{figure}

Tables~\ref{tab_augsburg_results} and \ref{tab_houston_results} report class-wise recall together with OA, AA, and Kappa. On Augsburg, CRFNet achieves the highest OA of 92.04\%. Its strongest class-wise results are obtained for low plants and allotment. The corresponding classification maps in Fig. \ref{fig:augsburg-res} further verify these results. In the enlarged region, CRFNet reproduces the main forest area and the adjacent low-plants region in the ground truth, with little false labeling. It is demonstrated that our CRFNet preserves local land-cover patterns while distinguishing vegetation classes. 

\begin{table}[!t]
\centering
\caption{Classification performance of different methods on the Augsburg dataset. The bold and underline denote the best and second results.}
\label{tab_augsburg_results}
\setlength{\tabcolsep}{2pt}
\resizebox{0.95\linewidth}{!}{
\begin{tabular}{c|ccccccc>{\columncolor{CRFNetShade}}c}
\toprule
Class & S$^2$ENet & DFINet & AsyFF & ExViT & HCT & MACN & RSCNet & CRFNet \\
\midrule
Forest & \textbf{98.10} & 97.38 & 97.56 & 90.04 & 97.26 & \underline{97.84} & 96.07 & 96.77 \\
Residential & \underline{99.08} & 98.37 & \textbf{99.16} & 95.44 & 98.63 & 98.73 & 96.40 & 97.94 \\
Industrial & 12.19 & 61.31 & \underline{61.38} & 34.58 & 36.61 & 43.94 & \textbf{78.38} & 56.48 \\
Low plants & 91.78 & 92.63 & 83.47 & 90.68 & 93.54 & \underline{94.11} & 94.08 & \textbf{96.62} \\
Allotment & 45.12 & 49.33 & 42.64 & 51.82 & 51.43 & 52.39 & \underline{58.32} & \textbf{61.38} \\
Commercial & 1.22 & 3.54 & 5.58 & \textbf{28.63} & \underline{7.97} & 6.80 & 1.96 & 3.24 \\
Water & 24.09 & 26.61 & 45.81 & 17.65 & 46.34 & 48.54 & \textbf{50.07} & \underline{48.91} \\
\midrule
OA & 88.22 & 90.66 & 88.25 & 86.65 & 90.33 & 91.06 & \underline{91.50} & \textbf{92.04} \\
AA & 53.08 & 61.31 & 62.23 & 58.41 & 61.68 & 63.19 & \textbf{67.89} & \underline{65.90} \\
Kappa & 86.47 & 86.47 & 83.13 & 80.79 & 86.03 & 87.06 & \underline{87.76} & \textbf{88.43} \\
\bottomrule
\end{tabular}
}
\end{table}

\begin{table}[!t]
\centering
\caption{Classification performance of different methods on the Houston2013 dataset. The bold and underline denote the best and second results.}
\label{tab_houston_results}
\setlength{\tabcolsep}{2pt}
\resizebox{0.95\linewidth}{!}{
\begin{tabular}{c|ccccccc>{\columncolor{CRFNetShade}}c}
\toprule
Class & S$^2$ENet & DFINet & AsyFF & ExViT & HCT & MACN & RSCNet & CRFNet \\
\midrule
Healthy grass & 81.48 & \textbf{82.91} & 82.72 & 82.72 & 81.10 & \underline{82.90} & 82.53 & 82.81 \\
Stressed grass & 84.68 & 85.15 & 83.93 & 85.15 & 86.09 & \textbf{100.0} & 79.98 & \underline{94.83} \\
Synthetic grass & 99.01 & 99.60 & \textbf{100.0} & 99.21 & \underline{99.80} & \underline{99.80} & 99.60 & \textbf{100.0} \\
Trees & 92.14 & 92.33 & 91.95 & 91.95 & 98.01 & \underline{98.11} & \textbf{99.15} & 97.16 \\
Soil & \textbf{100.0} & \underline{99.81} & \textbf{100.0} & \textbf{100.0} & \textbf{100.0} & \textbf{100.0} & 99.23 & \textbf{100.0} \\
Water & \textbf{100.0} & \textbf{100.0} & 95.80 & \underline{98.60} & 95.10 & 95.80 & \underline{98.60} & 95.80 \\
Residential & 89.93 & \underline{93.47} & \textbf{96.18} & 86.66 & 83.49 & 78.17 & 86.57 & 88.43 \\
Commercial & 90.18 & 82.58 & 81.14 & \underline{91.05} & 86.43 & 80.46 & 87.97 & \textbf{94.02} \\
Road & \underline{92.63} & 89.05 & 86.12 & 91.12 & 86.69 & 90.93 & 90.93 & \textbf{92.82} \\
Highway & 64.86 & 56.56 & 64.58 & 64.86 & 79.05 & 55.88 & \textbf{96.14} & \underline{89.38} \\
Railway & \underline{97.60} & 94.15 & 86.76 & 77.06 & 95.97 & \textbf{99.33} & \textbf{99.33} & 93.26 \\
Parking lot 1 & 88.09 & 94.72 & 90.59 & 88.76 & 98.46 & \textbf{99.71} & 96.93 & \underline{98.75} \\
Parking lot 2 & \underline{91.93} & 89.12 & \textbf{92.28} & 87.72 & 90.88 & 88.77 & 89.47 & 88.77 \\
Tennis court & \textbf{100.0} & \textbf{100.0} & 95.14 & \underline{97.17} & \textbf{100.0} & \textbf{100.0} & \textbf{100.0} & \textbf{100.0} \\
Running track & \textbf{100.0} & \textbf{100.0} & \textbf{100.0} & 98.52 & \textbf{100.0} & \textbf{100.0} & \textbf{100.0} & \underline{99.79} \\
\midrule
OA & 89.56 & 88.59 & 87.95 & 87.42 & 90.66 & 89.81 & \underline{92.66} & \textbf{93.75} \\
AA & 91.50 & 90.63 & 89.81 & 89.37 & 92.07 & 91.33 & \underline{93.76} & \textbf{94.39} \\
Kappa & 88.69 & 87.64 & 86.96 & 86.38 & 89.86 & 88.93 & \underline{92.04} & \textbf{93.22} \\
\bottomrule
\end{tabular}
}
\end{table}

On Houston2013, CRFNet obtains the highest OA, AA, and Kappa of 93.75\%, 94.39\%, and 93.22\%, respectively. It ranks first or second in nine of the fifteen classes. In terms of class-wise accuracy, CRFNet achieves the best results for Commercial and Road, reaching 94.02\% and 92.82\%, respectively, and attains 100\% accuracy for Synthetic grass, Soil, and Tennis court. These results demonstrate the strong overall classification performance of our CRFNet on this dataset.

\begin{table}[!t]
\centering
\caption{Comparison of model complexity on the Augsburg dataset.}
\label{tab_complexity}
\setlength{\tabcolsep}{2pt}
\renewcommand{\arraystretch}{0.96}
\resizebox{0.95\linewidth}{!}{%
\begin{tabular}{l|ccccccc>{\columncolor{CRFNetShade}}c}
\toprule
Metric & S$^2$ENet & DFINet & AsyFF & ExViT & HCT & MACN & RSCNet & CRFNet \\
\midrule
Params (M) & 1.7055 & 0.8460 & 2.0987 & 1.8807 & 4.7811 & 2.0199 & 2.8812 & 0.3492 \\
FLOPs (G) & 0.2065 & 0.12464 & 0.1789 & 0.2897 & 0.0437 & 0.1672 & 0.3346 & 0.0821 \\
\bottomrule
\end{tabular}}
\end{table}

\subsection{Computational Complexity}

Table \ref{tab_complexity} compares the parameter counts and FLOPs of different methods on the Augsburg dataset. CRFNet requires only 0.3492 M parameters, achieving the smallest model size among all compared methods. Specifically, it reduces the parameter count by 58.72\% and 87.88\% compared with DFINet and RSCNet, respectively. In terms of computational complexity, CRFNet requires 0.0821 GFLOPs, ranking second only to HCT. These results demonstrate that CRFNet combines a compact model size with low computational cost.

\subsection{Ablation Study}

Table \ref{tab_ablation} presents the ablation results of CLRA and FCF on both datasets. Without either component, the baseline achieves OA values of 90.71\% and 89.36\%, respectively. Introducing CLRA alone yields gains of 0.53 and 2.67 percentage points. Similarly, incorporating FCF alone increases OA to 91.51\% and 92.24\%. Combining both components achieves the best OA values. These results demonstrate that both components contribute to classification performance and that their combination provides further improvements on both datasets.

\begin{table}[!t]
\centering
\caption{Ablation experiments. OA (\%) is reported.}
\label{tab_ablation}
\renewcommand{\arraystretch}{0.9}
\begin{tabular}{cc|cc}
\toprule
CLRA & FCF & Augsburg & Houston2013 \\
\midrule
$\times$ & $\times$ & 90.71 & 89.36 \\
$\times$ & $\checkmark$ & 91.51 & 92.24 \\
$\checkmark$ & $\times$ & 91.24 & 92.03 \\
\rowcolor{CRFNetShade}$\checkmark$ & $\checkmark$ & \textbf{92.04} & \textbf{93.75} \\
\bottomrule
\end{tabular}
\end{table}

\section{Conclusion}

We proposed CRFNet, a lightweight dual-branch network for multi-source remote sensing image classification. Specifically, CLRA progressively enhances discriminative spatial dependencies through cross-layer attention refinement and top relation selection, while FCF exploits cross-modal complementary information through bidirectional frequency-domain interactions. Extensive experiments on two benchmark datasets demonstrate that the proposed CRFNet consistently outperforms state-of-the-art methods.

\bibliographystyle{IEEEtran}
\bibliography{ref}

\end{document}